\documentclass[a4paper,aps,prl,twocolumn,groupedaddress,showkeys,showpacs,floats,floatats,floatfix]{revtex4-1}
\usepackage[utf8]{inputenc}
\usepackage[english]{babel}
\usepackage{graphicx}
\usepackage{dcolumn} 
\usepackage{bm}
\usepackage{natbib}
\usepackage{latexsym}
\usepackage{mathrsfs}
\usepackage{amssymb}
\usepackage{amsmath}
\usepackage{amscd}
\usepackage{color}
\usepackage{pifont}
\usepackage{pstricks,pst-node,pst-text,pst-3d}
\usepackage{verbatim}
\usepackage[T1]{fontenc}
\newrgbcolor{Red}{1.0 0.0 1.0}
\begin{document}
\title{Ratchet transport and periodic structures in parameter space}
\author{A.~Celestino$^1$, 
C.~Manchein$^2$, H.A.~Albuquerque$^1$ and M.W.~Beims$^{2}$}
\affiliation{$^1$Departamento de F\'\i sica, Universidade do Estado de
  Santa Catarina, 89219-710 Joinville, SC, Brazil} 
\affiliation{$^2$Departamento de F\'\i sica, Universidade Federal do Paran\'a,
         81531-980 Curitiba, PR, Brazil}
\date{\today}
%
\begin{abstract}
Ratchet models are prominent candidates to describe the transport phenomenum 
in nature in the absence of external bias. This work analyzes the parameter 
space of a discrete ratchet model and gives direct connections 
between chaotic domains and a family of {\it isoperiodic stable structures} 
with the ratchet current.  The isoperiodic structures appear along preferred 
direction in the parameter space giving a guide to follow the current, which 
usually increases inside the structures but is independent of the 
corresponding period. One of such structures has the shrimp-shaped form 
which is known to be an universal structure in the parameter space of 
dissipative systems. Currents in parameter space provide 
a direct measure of the momentum asymmetry of the multistable 
and chaotic attractors {\it times} the size of the corresponding basin of 
attraction. Transport structures are 
shown to exist in the parameter space of the Langevin equation 
with an external oscillating force.
\end{abstract}
%
\maketitle

The description of the ratchet transport of particles in nature has become an
actual and large studied problem due to the possibility to obtain transport 
properties without external bias. To obtain ratchet transport, spatiotemporal 
symmetries must be broken in the system \cite{artuso07}. Ratchets have become 
natural candidates to describe transport phenomena in Brownian 
\cite{astumian02,reimann} and molecular motors \cite{julicher97}, cold atoms 
\cite{dima05}, migration of bacteria 
\cite{lambert10}, cell mobility in cancer metastasis 
\cite{campbell09}, granular gas \cite{meer10}, fluid transport 
\cite{hanggiFLUID08}
and in more general areas like classical and quantum physics
\cite{kohler}, chemistry \cite{gong,ratchets7} and biophysics
\cite{dittrich10}.
These are just some references in the distinct areas, since the actual 
literature related to ratchets is enormous.

A common feature of interest in all areas of ratchets applications is the
understanding, achievement and control of transport. A priori, dynamical 
variables and parameters of the system (like temperature, dissipation, noise
intensity, external forces etc.), which control the dynamics, are deeply 
interconnected so that it is very hard to make general statements about the 
ratchet current (${\cal RC}$)
as a function of the parameters. The precise determination of the nature of 
transport in unbounded systems is still not fully understood, thus it is very 
desirable to achieve and/or recognize ``patterns'' or ``structures'' in the 
parameter space which are directly connected to transport properties.
Even more attractive if such transport structures present universal features 
observed in a large class of dynamical systems.

This Letter analyzes the parameter space of a ratchet model and shows the
relation between ratchet currents with a family of {\it isoperiodic stable 
structures} (ISSs) and {\it chaotic domains} in parameter space. In this way a 
remarkable complete connection between parameters of the system and the
${\cal RC}$ is given, and therefore { general} clues for the origin 
of directed transport. To mention an example, one of the 
ISSs observed here, which has the shrimp-shaped form [see
Fig.~\ref{zoom}(c)], has already appeared in the parameter space of generic 
dynamical systems and applications. Such shrimps were found to be universal 
structures in the parameter space of dissipative systems like { 
maps \cite{jasonPRL93,grebogi93,murilo96} and continuous models 
\cite{bonattoR07}, among others.}
Very recently their were also observed in experiments 
with electronic circuits \cite{stoop10}. We show here that such shrimp-shaped 
structures are also essential concerning directed 
transport in nature and we stress that this is valid for any ratchet model 
and applications.

In order to show generic properties of the ${\cal RC}$ in 
the parameter space, we use a Map which presents all
essential features regarding unbiased current \cite{casatiPRL07}

\begin{eqnarray}
  M:\left\{
\begin{array}{ll}
  p_{n+1} = \gamma p_n + K[\sin(x_n)+a\sin(2x_n+\phi)], \\
  x_{n+1} = x_n + p_{n+1},
\end{array}
\right.
\label{map}
\end{eqnarray}
where $p_n$ is the momentum variable conjugated to $x_n$, $n=1,2,\ldots,N$
represents the discrete time and $K$ is the nonlinearity parameter. The 
dissipation parameter $\gamma$ reaches the overdamping limit for $\gamma=0$
and the conservative limit for $\gamma=1$.
The ratchet effect appears due to the spatial asymmetry, which occurs with 
$a\ne0$ and $\phi\ne m\pi$ ($m=1,2,\ldots$), in addition to the time reversal 
asymmetry for $\gamma\ne1$. { The ${\cal RC}$ of the above 
model was studied \cite{casatiPRL07} for fixed $K=6.5$ in the dissipation 
interval $0\le\gamma<1$. It was shown that close to the limit $\gamma=1$ 
the ${\cal RC}$ arises due to the mixture of chaotic motion with tiny 
island (accelerator modes) from the conservative case, while for smaller
values of $\gamma$, chaotic and stable periodic motion generates the current.

\begin{figure}[htb]
  \centering
  \includegraphics*[width=0.94\columnwidth]{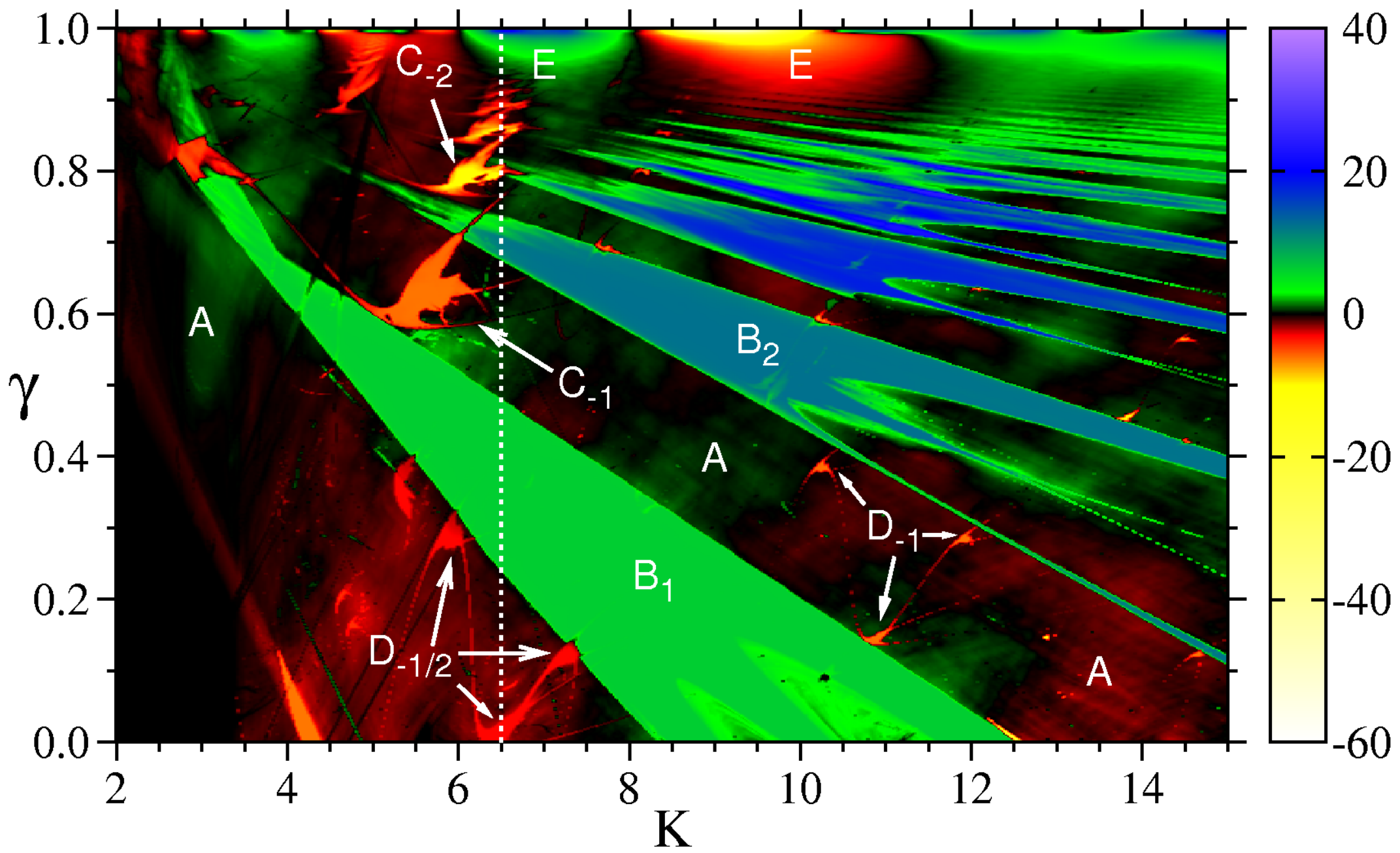}
  \caption{(Color online) The current (see color bar) plotted in the parameter 
 space ($K,\gamma$) with a   grid of $600\times600$ points, $a=0.5$, 
$\phi=\pi/2$,  $10^5$ initial conditions with $<p_0>=<x_0>=0$ inside the
unit cell ($-2\pi,2\pi$), and $N=10^4$ iterations.}
  \label{PSS1}
\end{figure}
Figure \ref{PSS1} shows the ${\cal RC}$  (colors) as a function of the 
dissipation parameter $\gamma$ and the nonlinearity parameter $K$.
A remarkable complex structure of colors
is evident, where each color is related to a given value of the current (see 
color bar). Black colors are related to close to zero currents; green, blue 
to purple colors are related to increasing positive currents while red, 
yellow to white colors related to increasing negative currents. The white
straight line at $K=6.5$ corresponds to the case analyzed recently 
\cite{casatiPRL07}. 
Three main regions with distinct behaviors can be identified: (i) a large 
{\it ``cloudy'' background}, identified as $A$ in Fig.~\ref{PSS1}, mixed 
with black, red and green colors, showing a mixture of zero, small negative 
and positive currents, respectively; (ii) several
{\it structures with sharp borders} and distinct colors, which are embedded 
in the cloudy background region and are identified in Fig.~\ref{PSS1}
as $B_L, C_L$ and $D_L$ { ($L$ is an integer or rational number)}; 
(iii) strong positive and negative currents (region
$E$), with not well defined borders which occur close to the conservative limit 
$\gamma=1$. At next we explain in more details these distinct regions 
by analyzing other quantities.

Figure \ref{period} shows the parameter space ($K,\gamma$) for the period-$q$ 
from the orbits. Periodic { stable} motion is restricted to well defined 
structures while the black background is related to the chaotic motion. This 
was checked by determining (not shown) the parameter 
space for the largest Lyapunov exponent (LE). Zero and negative LEs are related
to the periodic motion and positive LEs to the black regions of
Fig.~\ref{period}. This already allow us to associate the cloudy region $A$ 
from Fig.~\ref{PSS1} with the chaotic motion. Thus the small portions of 
negative/positive currents (red and green clouds) are due to the chaotic 
transport, consequence of the asymmetry of the chaotic attractor \cite{dima05}.
\begin{figure}[htb]
  \centering
  \includegraphics*[width=0.84\columnwidth]{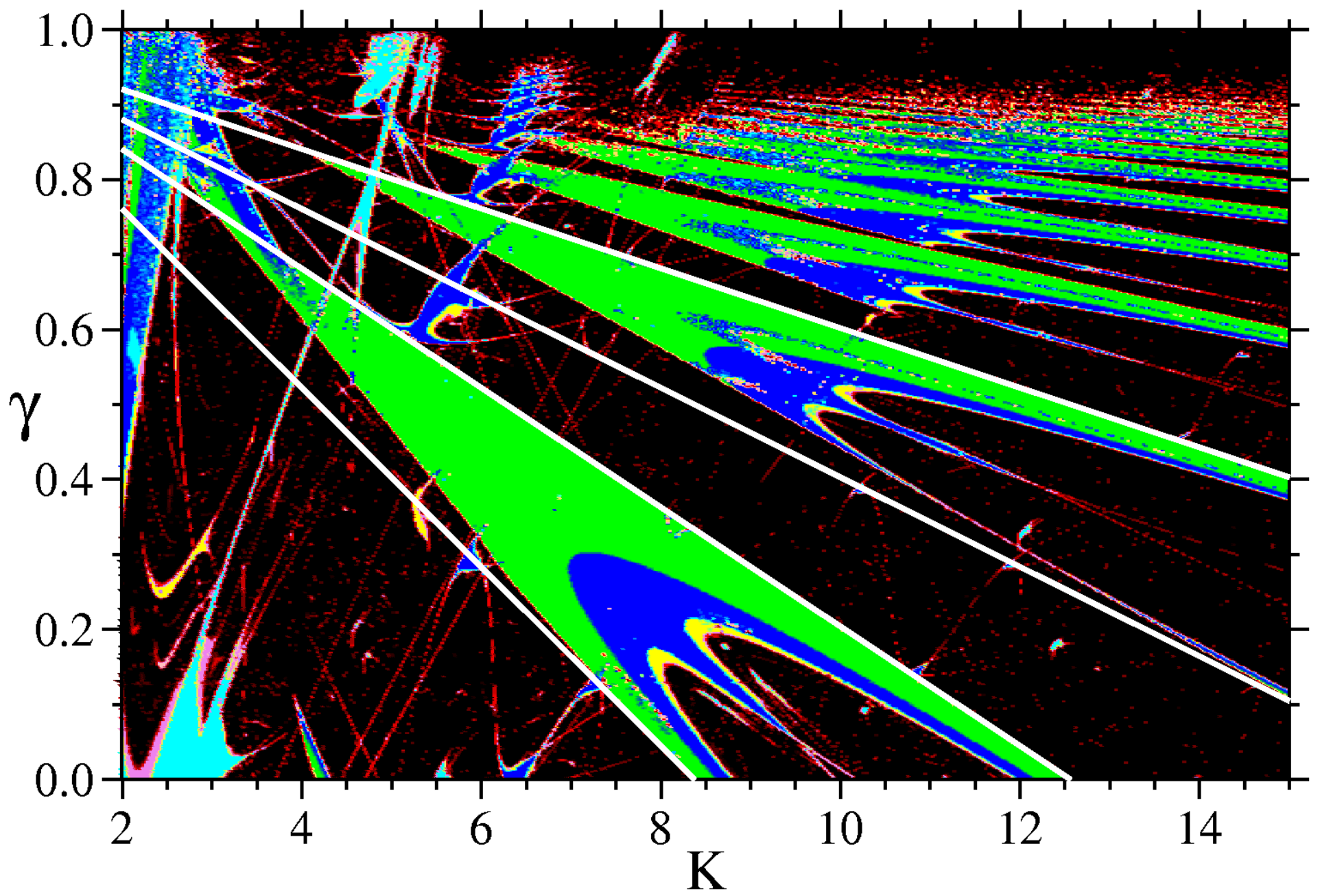}
 \caption{(Color online) Period-$q$ values in parameter space ($K,\gamma$): 
green: $q = 1$, blue: $q = 2$, cyan: $q = 3$, yellow: $q = 4$, pink: $q=6$,
red $q\ge 8$ and black 
for no period. In this case only one initial condition is used 
($x_0=0.5,p_0=0.3$), $N=10^6$ iterations and a grid of
$600\times600$ points. } 
  \label{period}
\end{figure}

{ Map (\ref{map}) is periodic in $x$ with period $2\pi$. Thus 
the condition for a period-$q$ orbit is $\sum_{i=n}^{n+q-1} p_i= 2\pi m$, 
with integer $m=\ldots,-2,-1,0,1,2,\ldots$ which represent the net number 
of left (negative) or right (positive) $2\pi$ jumps in $x$ over the period. 
The mean momentum for 
a given period-$q$ is then $\overline p_q=\sum_{i=n}^{n+q-1} p_i/q= 2\pi m/q$ 
which can be written as $2\pi L$, so that $L=m/q$. Therefore $L$, which 
gives $\overline p_q$ in units of $2\pi$, can assume fractional 
values (positive or negative).}
All sharp borders from $B_L, C_L$ and $D_L$ in Fig.~\ref{PSS1}
coincide with those of Fig.~\ref{period} and a direct connection between 
periodic { stable} motion and the ${\cal RC}$ is given. 
Regarding these sharp structures we observe three different 
ISSs: $B_L$ composed by the sequence 
$L=1,2,\ldots$ of dominant diagonal structures with main period $q=1$
(see green main regions in Fig.~\ref{period}) which extend  themselves along 
a large range of $K$ values. As $K$ increases and $\gamma$ decreases, inside 
each structure $B_L$ we observe $1 \times 2^n$ (green $\to$ blue $\to$ yellow  
$\to \ldots$) and $3 \times 2^n$ (cyan $\to$ pink  $\to \ldots$) doubling 
cascades 
bifurcations. Comparing to Fig.~\ref{PSS1}, we observe that the current
inside each structure $B_L$ is independent of the period $q$ and thus, on the 
bifurcation points. At the doubling bifurcation cascades 
$1\times 2^ n$ we observed that $m\to 2m$ so that $L$ 
($=m/q=\overline p_q/2\pi$) remains constant inside each structure $B_L$. On 
the other hand, even though different isoperiodic $B_L$ structures have the 
same period, ${\cal RC}$s increase with $L$, as can be seen in Fig.~\ref{PSS1} 
when $\gamma$ and $L$ increase. { Structures
$B_L$ are very similar to the cuspidal singularities \cite{jason06}}. 
Since the boundary of 
the dominant structures $B_L$ are born at period $q=1$, these borders can be 
calculated analytically from the eigenvalues of the Jacobian of the map 
(\ref{map}) after one iteration. Using $\phi=\pi/2$, the fixed points from 
(\ref{map}) can be calculated from
$p^{(1)} = 2\pi L$ ($L$ integer) and
$2\pi L (\gamma-1) + K\left[\sin{(x^{(1)})}+ a\cos{(2x^{(1)})}\right]=0$.
The solutions are $x^{(1)}_j=\arctan{(\alpha^{(-)},\pm \beta^{(+)})}$ ($j=1,2$) 
and  $x^{(1)}_s=\arctan{(\alpha^{(+)},\pm \beta^{(-)})}$ ($s=3,4$)
where $\alpha^{(\mp)}=\mp\sqrt{K\left[3K+8L\pi (\gamma-1) \right]}+K$ and
$\beta^{(\pm)}=\sqrt{8KL\pi(1-\gamma)\pm 2\sqrt{K^2\left[3K+8L\pi (\gamma-1) \right]}}$.
Substituting these solutions in the Jacobian of the map (\ref{map}) we
obtain analytical expressions $\lambda(\gamma,K,L)$ for the two eigenvalues.
When $\lambda(\gamma,K,L)=+1$ (born of period-$1$), we obtain a relation 
between $\gamma,K$ and $L$ where period-$1$ orbits are born in parameter 
space. The solutions for $\lambda(\gamma,K,L)-1=0$, for all fixed
points $x^{(1)}_j$, are $\gamma_1^{(L)}=1-3K/(8\pi L)$ (lower border) and  
$\gamma_2^{(L)}=1-K/(4\pi L)$ (upper border). Both curves define exactly, for 
a given $L$, the sharp period-$1$ $B_L$ boundaries of Figs.~\ref{PSS1} and 
\ref{period}. The border of the first large dominant structure $B_1$ is 
obtained from $L=1$, the second one $B_2$ from $L=2$, and so on. {
See white lines $\gamma_1^{(L=1)},\gamma_2^{(L=1)},\gamma_1^{(L=2)},$ and 
$\gamma_2^{(L=2)}$ in Fig.\ref{period}}. Looking carefully however, all (for any 
$L$) lower borders $\gamma_2^{(L)}$ do not match exactly with the simulations. 
For these regions the basin of attraction from the chaotic attractor is much 
larger compared to the basin of attraction related to the fixed point, and 
thus the {current related to} period $1$ is too small and 
is not observed. This is also exactly what is observed in Fig.~3 from 
\cite{casatiPRL07}, where the left limits of the $L$ intervals are inside the 
chaotic region. As $L$ increases, 
$\gamma_1^{(L\to\infty)}\to\gamma_2^{(L\to\infty)}\to1$ and the 
dominant structures approach to each other more and more.

The second kind of relevant ISSs, $C_L$, can be
visualized in Fig.~\ref{PSS1} close to $K=6.0$ and $\gamma>0.6$. They also 
are ordered in a sequence of structures $L=-1,-2,\ldots$ which approach each 
other as $\gamma$ increases, defining a direction in the parameter space 
obtained by the straight line $\gamma= 0.2845K-0.994925$, { along 
which negative currents} increase [see dashed line in Fig.~\ref{zoom}(a)]. 
The main period of each $C_L$ is $q=2$ (blue) [Fig.~\ref{zoom}(b)] but, 
inside each structure, 
$1 \times 2^n$ (blue $\to$ yellow $\to$ red $\to \ldots$) doubling cascades 
bifurcations appears when going 
to the border of the structures, { where the chaotic region is 
reached.} For clarification Figs.~\ref{zoom}(a)-(b) present the current 
and period as a magnification for { these} structure. Again we 
observe that current increases {(in modulus)} along the sequences as 
$\gamma$ increases.

The last observed ISSs, $D_L$, appear embedded in the cloudy chaotic 
background and present the { well known} shrimp-shaped form. For 
example, three { connected} structures 
($D_{-1}$) appear in red in the interval $10<K<12$ and $0.1<\gamma<0.4$ 
[see also Fig.~\ref{zoom}(c)]. { Another example of 
shrimp-shaped structures is demarked in Fig.~\ref{PSS1} by $D_{-1/2}$. } 
These structures have a main body with period $q=2$ and a succession of 
domains related to period-doubling route to chaos. Shrimps-shaped structures 
are also distributed in sequences along preferred direction in the parameter 
space, as can be seen by looking carefully to Fig.~\ref{PSS1}, where many 
shrimp-shaped structures are hidden behind the dominant $B_L$ structures.
{ Such preferential directions in parameter space appear to be
general properties of shrimp-shaped structures (see \cite{jasonPRL93}).
These} structures are abundant in the parameter
space as can be seen by the magnification shown in Figs.~\ref{zoom}(c)-(d).
As the parameter space is searched further and further for finer domains, a 
large amount of distinct ISSs appear, usually well organized 
and sometimes even connected to each other [see the connected shrimp-shaped
ISSs in Fig.~\ref{zoom}(c)]. Connected shrimps have the same current value. 
Although almost all ISSs present finite ${\cal RC}$s, some ISSs present
zero currents inside. We mention one example which can nicely be observed in 
Fig.~\ref{zoom}(a) close to $K=4.5$ and $\gamma=0.75$ [compare with 
the cyan ISS from Fig.~\ref{zoom}(b)]. Such very interesting 
cases will be analyzed in another work.
\begin{figure}[htb]
  \centering
  \includegraphics*[width=0.45\columnwidth]{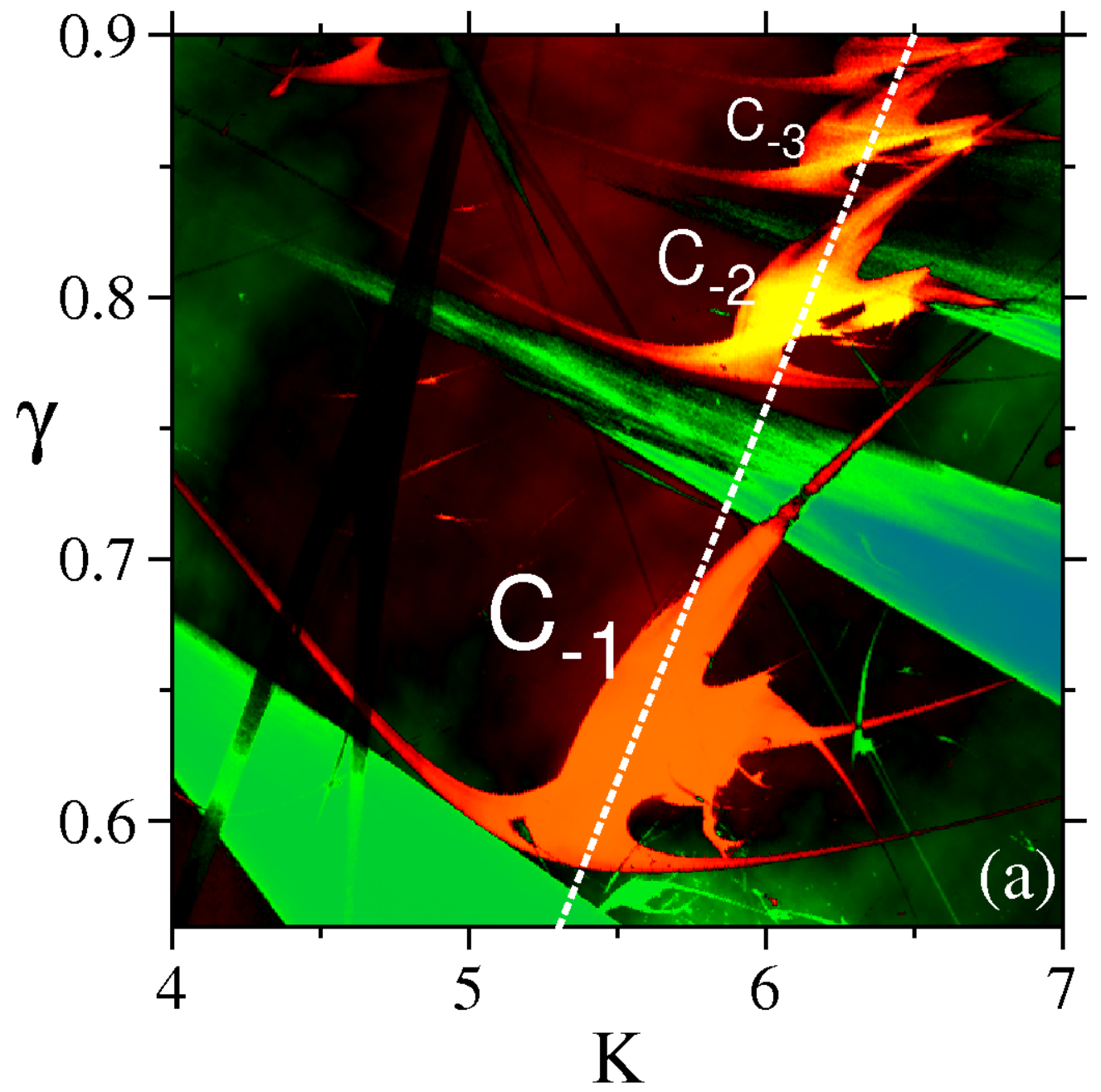}
  \includegraphics*[width=0.45\columnwidth]{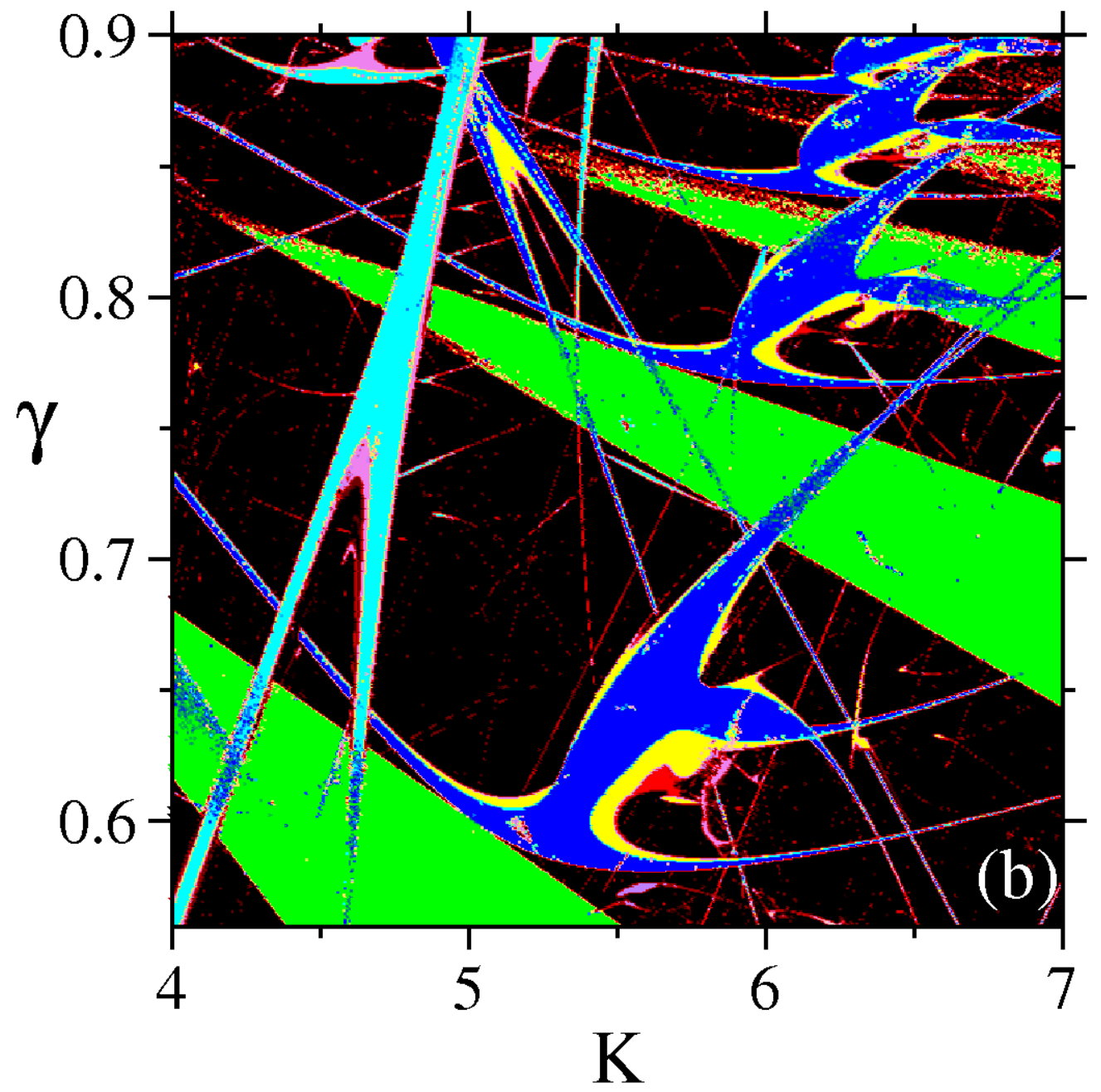}
  \includegraphics*[width=0.45\columnwidth]{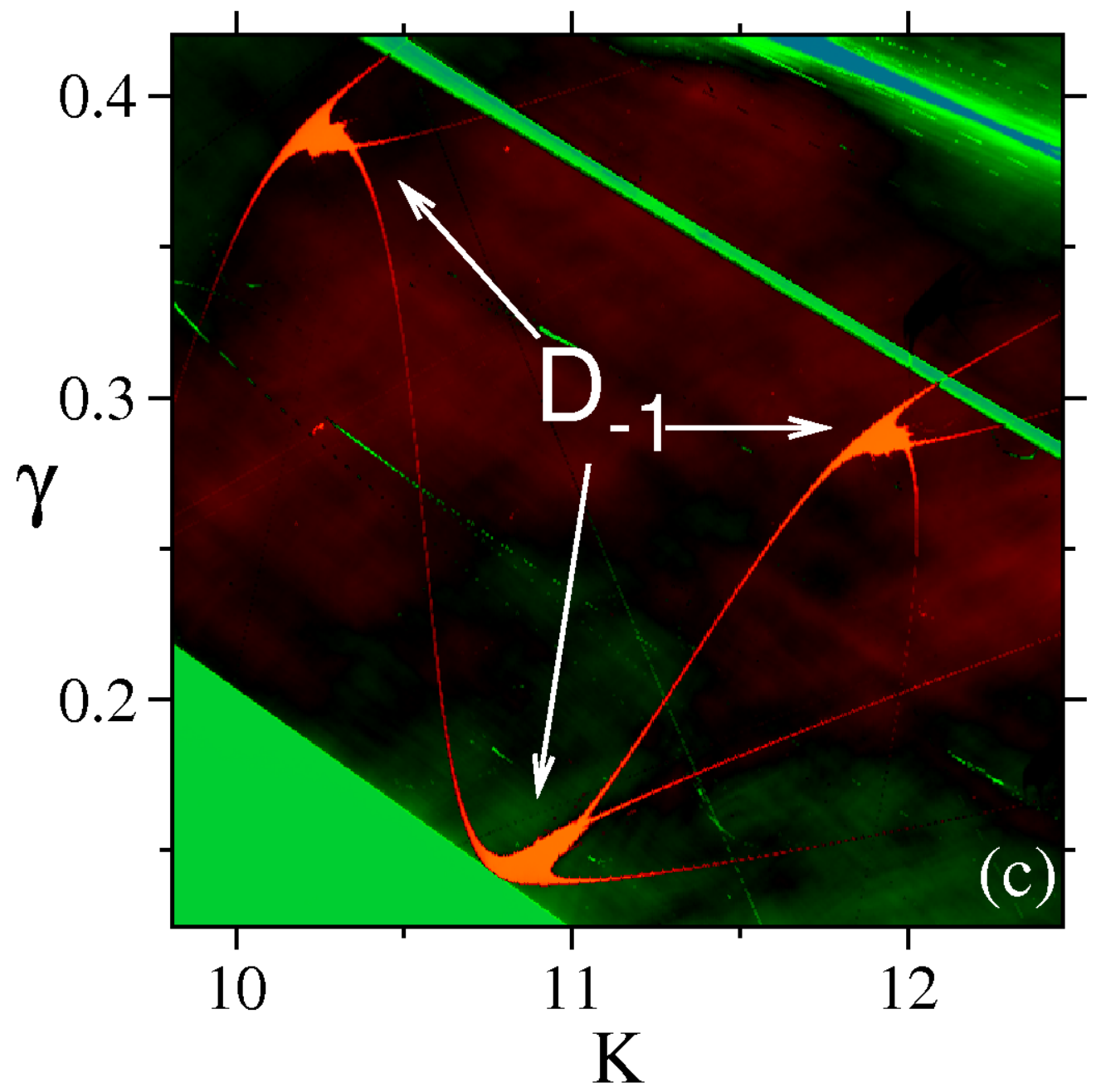}
  \includegraphics*[width=0.45\columnwidth]{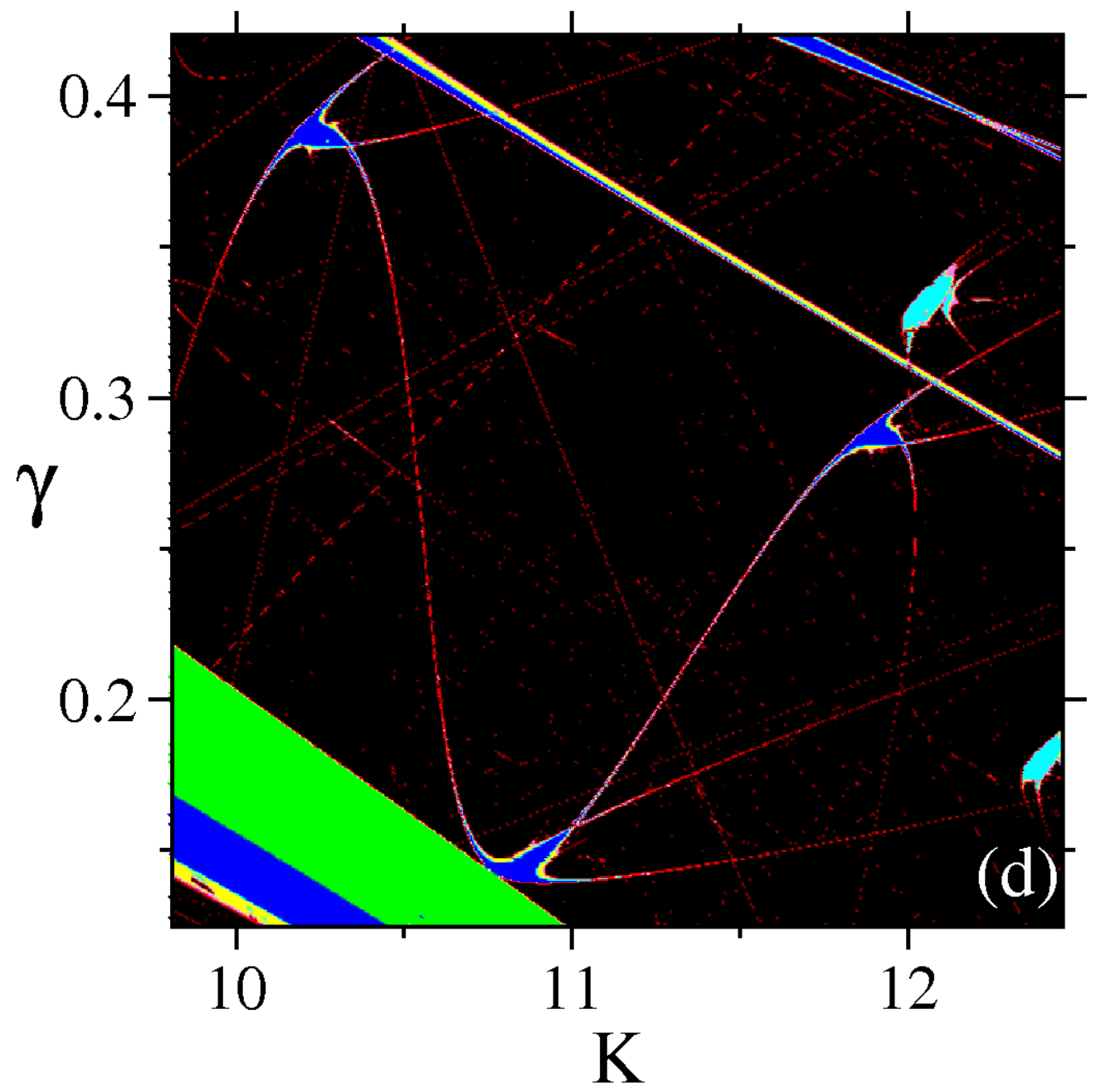}
  \caption{(Color online) Magnifications of Fig.~\ref{PSS1} (left) and 
 Fig.~\ref{period} (right) showing that the ISSs are abundant inside the 
chaotic region and usually organized 
along specific lines and sometimes connected to each other.}
  \label{zoom}
\end{figure}

The region $E$ presents larger currents close to $\gamma=1$ with not well 
defined borders. We start by mention
that the parameter space for the largest LE (not shown here) in this
region is positive and thus totally chaotic. We also clearly see from 
Figs.~\ref{PSS1} 
and \ref{period} that ISSs with larger $L$ start to overlap to each other 
when $\gamma\to1$ and thus the number of stable periodic points in phase 
space increases more and more. But the interesting point is that apparently 
the periodic structures are not directly responsible for the currents in 
region $E$. For one example see the 
preferential line along the structures $C_{-1}, C_{-2}\ldots$ in 
Fig.~\ref{zoom}(a), where the currents increase negatively as $\gamma\to1$. 
If the $C_L$ structures were responsible for the current in $E$, we would 
expect a very large negative current. However, this is not what happens since 
a current reversal occurs close to $\gamma\to1$ (see green region in 
Fig.~\ref{PSS1} at the end of the $C_{-1}, C_{-2}\ldots$ sequence). In fact, 
close to $\gamma=1$ we have regions of multistability and chaotic motion 
were the dynamics, which is a 
mixture of a large number of periodic and chaotic attractors, each one with 
his own basin of attraction, is very rich and complex. Such multistability 
regions were analyzed for the kicked rotor in the beautiful works of 
\cite{luciano08,feudel96}. The ``competition'' between multistable and 
chaotic attractors in order to generate the  ${\cal RC}$ is 
gained by the larger basin of attractions of chaotic attractors. Extensive 
numerical simulations show that close to $\gamma=1$ the 
unsharply borders do not change for larger iteration times and that the 
basins of attraction of periodic orbits are very small compared to the 
basins of chaotic attractors. This agrees with some recent works 
\cite{casatiPRL07,dima05,flach02, dittrich01} which suggest 
that the ${\cal RC}$ in this region is due to the mixture of chaotic motion 
with tiny island from the conservative case. In fact, accelerator 
modes from $\gamma=1$ are responsible for the asymmetry of chaotic attractors, 
generating the currents.

In order to show that ${\cal RC}$s present generic ISSs in the
parameter space of a more general class of dynamical systems, 
we analyze the zero temperature Langevin equation:
$\ddot x + \gamma\dot x - 5.0\left[\sin{(x)} 
+ 0.7\sin{(2x-\frac{\pi}{2})}\right]
 - K_t\sin{(1.0\,t)}=0$.
$K_t$ is the amplitude of the external time oscillating force,
$\gamma$ is the viscosity and the force coming from the ratchet
potential is identical from Eq.~(\ref{map}). 
\begin{figure}[htb]
  \centering
  \includegraphics*[width=0.9\columnwidth]{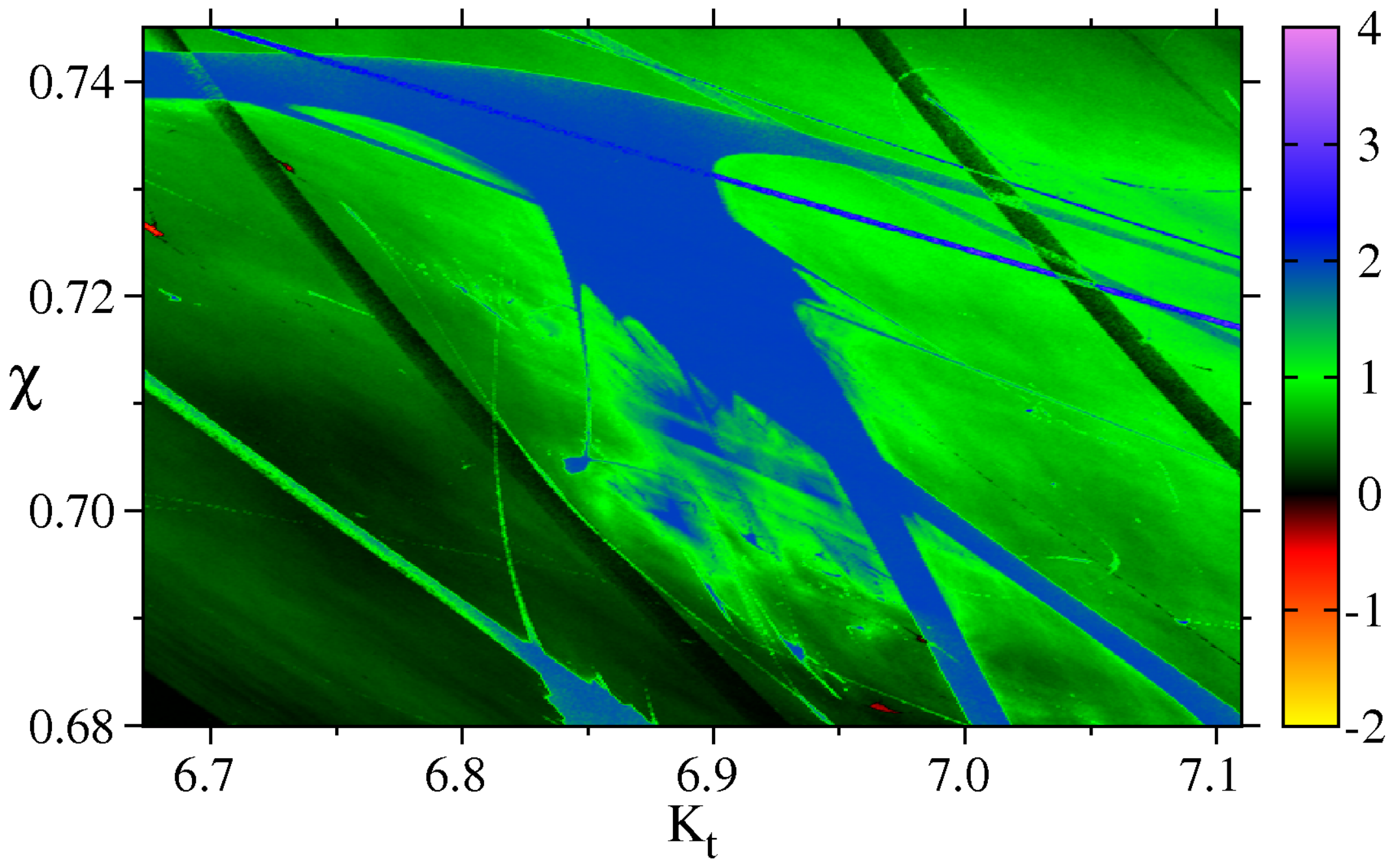}
  \caption{(Color online) Current in the parameter space 
($K_t,\chi=e^{(-\gamma)}$) for the Langevin equation.}
  \label{lang-ext}
\end{figure}
Figure \ref{lang-ext} shows the ${\cal RC}$ in a  smaller portion of the 
parameter space ($K_t,\chi=e^{(-\gamma)}$). As in Fig.~\ref{PSS1},
larger currents can be observed in the large blue shrimp-shaped ISS with 
main period $q=2$. Again doubling bifurcations occur inside the ISS (not 
shown), as observed in Figs.~\ref{period} and \ref{zoom}(d) for the 
$D_{-1}$ and $D_{-1/2}$ ISSs. Behind the main shrimp-shaped ISS, a cloudy 
background is observed and also a large amount of smaller ISSs, which can 
be better resolved when the parameter space is searched for finer and finer 
domains. Small finite temperatures will slowly transform the 
sharp borders of the ISSs into unsharply borders and simultaneously enlarge 
them since their attractors are more stable under noise effects than the
chaotic attractors \cite{blackburn96}.

Concluding, universal ISSs in parameter space are shown to generate 
large ${\cal RC}$s and to
organize themselves along preferential directions, which allows to make 
predictions of the ${\cal RC}$s along the such directions.
Inside a given structure, ${\cal RC}$s 
are independent of the  period of the orbit. The essential property to 
obtain finite ${\cal RC}$s is the momentum asymmetry of attractors in 
phase space { {\it times} the size of the corresponding basin of
attraction and can be expressed as 
${\cal RC}=\sum_{i=1}^{N_a} \left<p\right>_i S_i$, 
where $N_a$ is the number of attractors, $\left<p\right>_i$ is the mean 
momentum of attractor $i$ and $S_i$ is the size (normalized) of attractor $i$}.
Thus Figs.~\ref{PSS1} and \ref{lang-ext} are a  
{\it direct quantitative measure} of ${\cal RC}$s. We stress that the
isoperiodic stable transport structures $B_L$ (cuspidal-shaped),
$D_L$ (shrimp-shaped) and $C_L$ are universal patterns which should appear in 
the parameter space of any dissipative ratchet system, independent of its
application in nature. This was ratified by showing the appearance of the
shrimp-shaped ISSs in the parameter space of the Langevin equation with an 
external unbiased field. It would be very likely to observe 
the ${\cal RC}$s ISSs in real experiments. 

\vspace*{-0.5cm}

%
\end{document}